\documentclass[aps,prb,twocolumn,superscriptaddress,showpacs,showkeys,floatfix]{revtex4}

\usepackage{graphicx,color}
\newcommand{\jwj}[1]{\textcolor{red}{#1}}
\graphicspath{{figs/}}
\begin{document}
\title{Mechanical Properties of Single-Layer Black Phosphorus}
\author{Jin-Wu Jiang}
    \altaffiliation{Corresponding author: jwjiang5918@hotmail.com}
    \affiliation{Shanghai Institute of Applied Mathematics and Mechanics, Shanghai Key Laboratory of Mechanics in Energy Engineering, Shanghai University, Shanghai 200072, People's Republic of China}
\author{Harold S. Park}
    \altaffiliation{Corresponding author: parkhs@bu.edu}
    \affiliation{Department of Mechanical Engineering, Boston University, Boston, Massachusetts 02215, USA}

%\date{22 December 2009}
\date{\today}
\begin{abstract}
The mechanical properties of single-layer black phosphrous under uniaxial deformation are investigated using first-principles calculations. Both Young's modulus and the ultimate strain are found to be highly anisotropic and nonlinear as a result of its quasi-two-dimensional puckered structure. Specifically, the in-plane Young's modulus is 41.3~{GPa} in the direction perpendicular to the pucker, and 106.4~{GPa} in the parallel direction. The \jwj{ideal} strain is 0.48 and 0.11 in the perpendicular and parallel directions, respectively.
\end{abstract}

\pacs{68.65.Ac, 62.25.-g}
\keywords{Black Phosphrous, Young's Modulus}
\maketitle
\pagebreak

%\section{Introduction}
Few-layer black phosphorus (BP) is another interesting quasi two-dimensional system that has recently been explored as an alternative electronic material to graphene, boron nitride, and the transition metal dichalcogenides for transistor applications\cite{LiL2014,LiuH2014,BuscemaM2014,Castellanos-GomezA2014arxiv}. This initial excitement surrounding BP is because unlike graphene, BP has a direct bandgap that is layer-dependent.  Furthermore, BP also exhibits a carrier mobility that is larger than MoS$_{2}$\cite{LiuH2014}. The van der Waals effect in bulk BP was discussed by Appalakondaiah et.al.\cite{AppalakondaiahS2012prb} First-principles calculations show that single-layer BP has a band gap around 0.8~{eV}, and the band gap decreases with increasing thickness.\cite{LiuH2014,DuY2010jap} For single-layer BP, the band gap can be manipulated via mechanical strain in the direction normal to the BP plane, where a semiconductor-metal transition was observed.\cite{RodinAS2014}

\begin{figure}[htpb]
  \begin{center}
    \scalebox{0.7}[0.7]{\includegraphics[width=8cm]{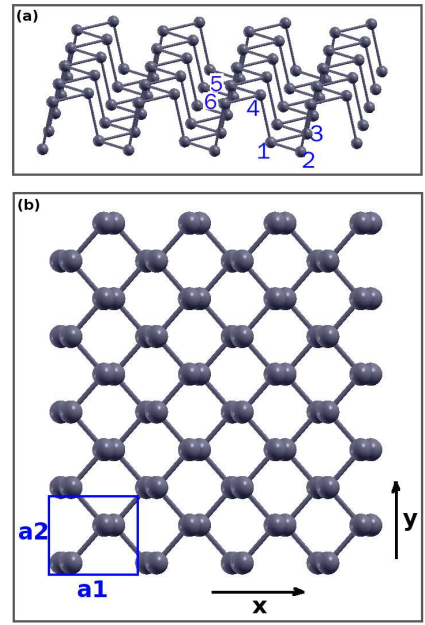}}
  \end{center}
  \caption{(Color online) Optimized configuration of single-layer BP. Top: perspective view illustrates the pucker along the y-direction. Bottom: top view of top image showing a square lattice structure.  Blue box represents the basic unit cell for single-layer BP.}
  \label{fig_cfg}
\end{figure}

\begin{figure*}[htpb]
  \begin{center}
    \scalebox{0.75}[0.75]{\includegraphics[width=\textwidth]{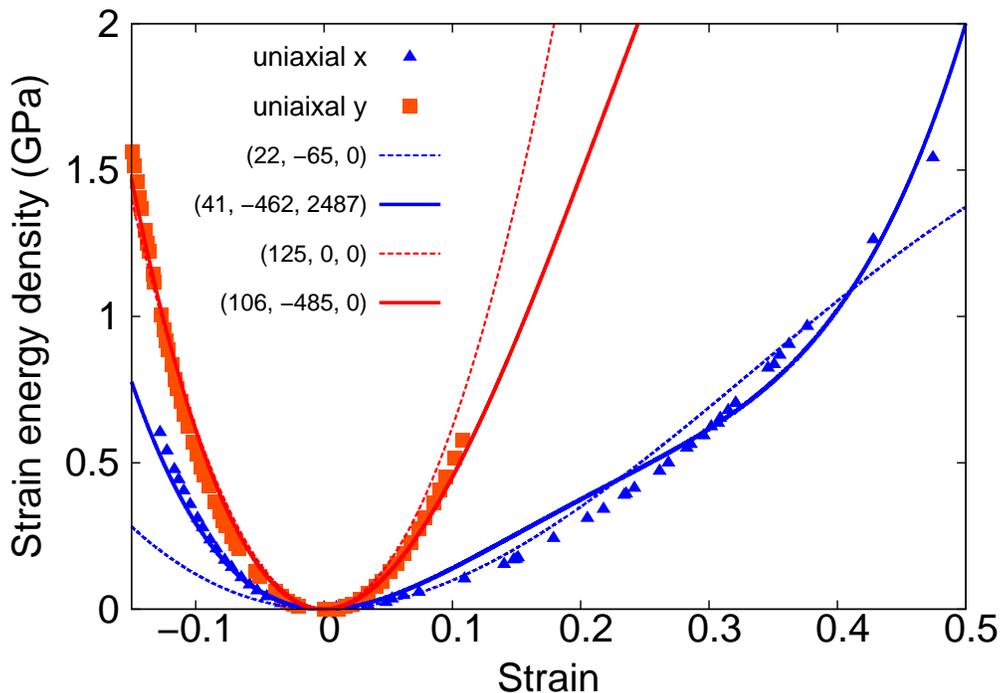}}
  \end{center}
  \caption{(Color online) Strain energy density for single-layer BP under uniaxial deformation in the x-direction (blue triangles), uniaxial deformation in the y-direction (red squares). Dashed lines represent the fitting to cubic function, i.e $y=Yx^2/2+C_{3}x^3/6$. Solid lines are fitting to the square function, i.e $y=Yx^2/2+C_{3}x^3/6+C_{4}x^4/24$. Fitting parameters $(Y, C_{3}, C_{4})$ are shown in the legend. Note that the cubic function gives poor fitting results, while the quartic function yields a good fitting. Right bottom insets show the side view of the single-layer BP under uniaxial compression and tension in the x-direction.}
  \label{fig_strain_energy}
\end{figure*}

Single-layer BP has a characteristic puckered structure as shown in Fig.~\ref{fig_cfg}~(a), which leads to two anisotropic in-plane directions.  The anisotropy engendered by the pucker should have a strong effect on the mechanical properties in the two orthogonal in-plane directions.  However, studies on the fundamental mechanical properties of single-layer BP are still lacking. 

In this letter, we report the highly anisotropic and nonlinear mechanical properties in single-layer BP that result from uniaxial deformation. Our first-principles calculations show that the Young's modulus is considerably smaller in the in-plane direction perpendicular to the pucker, while the single-layer BP is able to sustain a large mechanical strain of 0.48 in this perpendicular direction.

%\section{Structure and Simulation Details}
For the {\it ab initio} calculations, we used the SIESTA package\cite{SolerJM} to optimize the structure of single-layer BP. The local density approximation was applied to account for the exchange-correlation function with Perdew-Burke-Ernzerhof (PBE) parametrization\cite{PerdewJP1996prl} and the double-$\zeta$ basis set orbital was adopted. During the conjugate-gradient optimization, the maximum force on each atom is smaller than 0.01 eV\AA$^{-1}$. A mesh cut off of 120 Ry was used. Periodic boundary conditions were applied in the two in-plane transverse directions, while free boundary conditions were applied to the out-of-plane direction by introducing sufficient vacuum space of 15~{\AA}. Gamma point $k$ sampling was adopted for wave vector space.

Fig.~\ref{fig_cfg} shows the relaxed structure for a single-layer of BP of dimension $17.69\times16.74\times5.29$~{\AA} as visualized using XCRYSDEN.\cite{xcrysden}  The top panel is a perspective view that displays the puckered configuration of single-layer BP. In this puckered structure, each P atom is connected to three neighboring P atoms. There are two inequivalent P-P bonds in the relaxed structure, i.e $r_{12}=r_{13}=2.4244$~{\AA} and $r_{14}=2.3827$~{\AA}. Two inequivalent bond angles are $\theta_{213}=98.213^{\circ}$ and $\theta_{214}=\theta_{314}=97.640^{\circ}$. The blue box indicates the unit cell with four P atoms. The two lattice constants are $a_{1}=4.1319$~{\AA} and $a_{2}=3.6616$~{\AA}, and we chose 5.29~{\AA} to be the thickness of single-layer BP as it is the inter-layer spacing in bulk BP.\cite{DuY2010jap} These structural parameters are close to the experimental values.\cite{BrownA1965ac}  The top view shown in the bottom panel displays a square lattice structure for single-layer BP.  The Cartesian coordinates are set with the x-direction perpendicular to the pucker and the y-direction parallel with the pucker.

%\section{Results and discussion}
The puckered structure of single-layer BP implies that this material may be much more ductile in the x-direction than the y-direction, which should lead to strongly anisotropic mechanical properties. To illustrate this anisotropy, we stretch single-layer BP in two different ways; i.e uniaxial deformation in the x-direction and uniaxial deformation in the y-direction. The mechanical compression or tension is applied in the direction of the strain, while the strained structure is allowed to be fully relaxed, especially in the direction perpendicular to the strain, which is essential for the proper definition of Young's modulus.

Fig.~\ref{fig_strain_energy} shows the strain energy density, which is asymmetric with respect to compression or tension. This asymmetric behavior implies a strong nonlinear mechanical response in single-layer BP. We fit the strain energy density data to two different functions, i.e the cubic function $y=Yx^2/2+C_{3}x^3/6$ and the quartic function $y=Yx^2/2+C_{3}x^3/6+C_{4}x^4/24$, from which two key facts are gleaned.  First, the cubic function does not fit the simulation results well, which is in contrast to the quartic fitting, especially for the x-direction.  This fitting result discloses the important contribution from the quartic term $C_{4}$. In other words, single-layer BP responds highly nonlinearly during the mechanical deformation, especially in x-direction. The deformation is in such a large strain range that these nonlinear effects become important. Secondly, the Young's modulus in this material is anisotropic; i.e its value in the x direction ($Y=41.3$~{GPa}, $E=21.9$~{Nm$^{-1}$}) is less than half of that in the y direction (106.4~{GPa}, $E=56.3$~{Nm$^{-1}$}). The second value $E$ is the effective Young's modulus which is thickness independent. It is calculated by $E=Yh$, where $h=5.29$~{\AA} has been chosen as the thickness of the single-layer BP in the above calculations. Insets in the figure show the highly elastic of the single-layer BP in the x-direction. The pucker can be compressed or unfolded gradually upon external compression or tension in the x-direction. The pucker induced folding or unfolding mechanism is the origin for the smaller Young's modulus in the x-direction. The anisotropic Young's modulus arises from the anisotropically distributed electron wave functions in single-layer BP, which has been found to cause anisotropy for a range of physical properties.\cite{QiaoJarxiv14015045,XiaFarxiv14020270,WeiQarxiv14037882,QinGarxiv14060261} We note that Seifert et.al found the Young's modulus to be isotropic for BP in nanotube form based on a density functional tight binding approach.\cite{SeifertG2000cpl} This finding is in contrast to the results in present work, which is likely due to the different computational approaches that were utilized.

Owing to this puckered configuration, single-layer BP is highly ductile in the $x$ direction, and it can sustain strain as high as 0.48 in the $x$ direction. This \jwj{ideal} strain value is considerably higher than that in the $y$ direction of about 0.11. The \jwj{ideal} strain we discuss here is the strain at which single-layer BP fails at zero temperature. For strains above these values, fracture of the single-layer BP occurs. \jwj{It should be noted that the failure strain of the material would be lower at non-zero temperatures due to the the relative ease of instability nucleation when thermal effects are accounted for.}

The effective Young's modulus is independent of the thickness, so it is a proper quantity to compare the Young's modulus in the single-layer BP with other layered structures.The two-dimensional effective Young's modulus is 21.9~{Nm$^{-1}$} in the x direction and 56.3~{Nm$^{-1}$} in the y direction of the single-layer BP. These values are considerably smaller than the effective Young's modulus of the single-layer MoS$_{2}$, which is above 120.0~{Nm$^{-1}$}.\cite{CooperRC2013prb1,CooperRC2013prb2,BertolazziS,JiangJW2013sw}  The values are also one order of magnitude smaller than the effective Young's modulus in single-layer graphene, which is around 335.0~{Nm$^{-1}$}.\cite{LeeC2008sci,JiangJW2009young,JiangJW2010young}

%\section{Conclusion}
In conclusion, we have performed first-principles calculations to investigate the mechanical properties of single-layer BP.  We find that single-layer BP exhibits highly anisotropic and nonlinear mechanical properties due to its unique puckered structure.  Specifically, the in-plane Young's modulus in the direction perpendicular to the pucker is only half of that in the parallel direction, while the ultimate strain is much larger in the direction perpendicular to the pucker.

\textbf{Acknowledgements} We thank the anonymous referee for mentioning the actual meaning of the ideal strain. The work is supported by the Recruitment Program of Global Youth Experts of China and the start-up funding from Shanghai University. HSP acknowledges the support of the Mechanical Engineering department at Boston University.

%\bibliographystyle{aipnum4-1}
%\bibliography{biball}
%\bibliography{/home/JiangJinWu/Documents/papers/mypapers/latex/biball}
%merlin.mbs aipnum4-1.bst 2010-07-25 4.21a (PWD, AO, DPC) hacked
%Control: key (0)
%Control: author (8) initials jnrlst
%Control: editor formatted (1) identically to author
%Control: production of article title (-1) disabled
%Control: page (0) single
%Control: year (1) truncated
%Control: production of eprint (0) enabled
%
\end{document}